\documentclass{JHEP3} 
\usepackage{amsmath}

\title{$k$-string tensions in the 3-d
SU($N$) Georgi-Glashow model} 

\author{Dmitri Antonov
        \thanks{Permanent address: 
        ITEP, B. Cheremushkinskaya 25, RU-117 218, Moscow, Russia.}\\ 
	Dipartimento di Fisica ``E. Fermi'' dell'Universit\`a 
	di Pisa and I.N.F.N. sezione di Pisa \\
        Via Buonarroti 2, Ed. B-C, I-56127 Pisa, Italy \\ 
        E-mail: \email{antonov@df.unipi.it}}

\author{Luigi Del Debbio \\
	Dipartimento di Fisica ``E. Fermi'' dell'Universit\`a 
	di Pisa and I.N.F.N. sezione di Pisa\\
        Via Buonarroti 2, Ed. B-C, I-56127 Pisa, Italy \\ 
	E-mail: \email{ldd@df.unipi.it} 
} 

\abstract{ The classic argument by Polyakov showing that monopoles
produce confinement in the Higgs phase of the Georgi-Glashow model is
generalized to study the spectrum of $k$-strings. We find that the
leading-order low-density approximation yields Casimir scaling in the
weakly-coupled 3-d SU($N$) Georgi-Glashow model. Corrections to the
Casimir formula are considered. When $k\sim N$, the non-diluteness
effect is of the same order as the leading term, indicating that
non-diluteness can significantly change the Casimir-scaling
behavior. The correction produced by the propagating Higgs field is
also studied and found to increase, together with the non-diluteness
effect, the Casimir-scaling ratio. Furthermore, a correction due to
closed $k$-strings is also computed and is shown to yield the same
$k$-dependence as the one due to non-diluteness, but with the opposite
sign and a nontrivial $N$-dependence. Finally, we consider the
possible implications of our analysis for the SU($N$) analogue of
compact QED in four dimensions.}


\preprint{IFUP-TH 2003/43} 

\dedicated{This work is dedicated to the memory of Ian Kogan}

\begin{document}


\section{Introduction}

The current description of strong interactions is based on QCD, whose
dynamics becomes nonperturbative at energies of order
$\Lambda_\mathrm{QCD}$. Confinement, which can be defined as the
absence in the observed spectrum of asymptotic states that carry a
color charge, is commonly associated to a nonperturbative, linearly
rising potential between color sources - for recent discussions on
confinement see e.g.~\cite{Greensite:2003bk,Dosch:2000va} and
references therein. Such a potential is indeed consistent with the
Regge trajectories observed in the hadronic spectrum and hints to a
picture where hadrons are made of quarks joined by flux tubes of
chromoelectric field. String theory was originally introduced as an
effective theory to describe the dynamics of these flux
tubes. Numerical simulations of gauge theories regulated on a lattice
are an effective tool to investigate nonperturbative properties from
first principles; they have confirmed the existence of such a linear
potential, have provided good evidence in favor of a bosonic string
description, and have studied in some details the structure of the
flux tubes.

While this picture is generally accepted nowadays, the mechanism which
is responsible for confinement is still under active debate. New
aspects of the mechanism of confinement may be evidentiated by
studying SU($N$) gauge theories for $N>3$, under the assumption that
all these theories share the same basic properties with corrections
that are organised in powers of $1/N$. These ideas have triggered a
recent interest in the spectrum of $k$-strings in SU($N$) gauge
theories. A $k$-string is defined as the confining flux tube between
sources in higher representations, carrying a charge $k$ with respect
to the center of the gauge group $Z_N$, i.e. representations with
nonvanishing $N$-ality. These sources can be seen as the superposition
of $k$ fundamental charges, and charge conjugation exchanges $k$- and
$(N-k)$-strings, so that non trivial $k$-strings exist only for
$N>3$~\footnote{We shall not consider in this work high-dimensional
representations that are screened by gluons and do not yield a genuine
asymptotic string tension.}; their string tensions $\sigma_k$ can be -
and should be - used to constrain mechanisms of
confinement~\cite{DelDebbio:2002yp,Greensite:2002yn}. Results for the
values of $\sigma_k$ can be obtained by various approaches. Early
results, based on dimensional reduction arguments, suggest the
so-called ``Casimir scaling'' hypothesis for the ratio of string
tensions~\cite{Ambjorn:1984mb}:

\begin{equation}
\label{r}
R(k,N)\equiv\frac{\sigma_k}{\sigma_1} = \frac{k(N-k)}{N-1} \equiv C(k,N)
\end{equation}
where $\sigma_1$ is the fundamental string tension. Recent studies in
supersymmetric Yang-Mills theories and M-theory suggest instead a ``Sine
scaling'' formula:
\begin{equation}
R(k,N)=\frac{\sin\left(k \pi/N\right)}{\sin\left(\pi/N\right)}.
\end{equation}
Corrections are expected to both formulae, but the form of such
corrections is unknown for the physically relevant case of a four
dimensional, nonsupersymmetric, SU($N$) gauge theory.

In the large-$N$ limit, where the interactions between flux tubes are
suppressed by powers of $1/N$, the lowest energy state of the system
should be made of $k$ fundamental flux tubes connecting the sources;
hence:
\begin{equation}
R(k,N) \stackrel{{k-{\rm fixed}}\atop{N\to \infty}}{\longrightarrow} k.
\end{equation}
Both the Casimir and the Sine scaling formulae satisfy this
constraint; they also remain invariant under the replacement $k \to
(N-k)$, which corresponds to the exchange of quarks with
antiquarks. However, it has been argued in Refs.~\cite{Armoni:2003ji,
Armoni:2003nz} that the correction to the large-$N$ behavior should
occur as a power series in $1/N^2$ rather than $1/N$.~\footnote{One
should study with some care whether the arguments presented in
Refs.~\cite{Armoni:2003ji, Armoni:2003nz} hold independently of the
space-time dimensionality.} Clearly such a behavior would exclude
Casimir scaling as an {\it exact}\ description of the $k$-string
spectrum.

Recent lattice calculations have provided new results for the spectrum
of $k$-strings both in three and four
dimensions~\cite{Lucini:2000qp,Lucini:2001ej,Lucini:2001nv,
DelDebbio:2001kz,DelDebbio:2001sj,DelDebbio:2003tk}. They all confirm
that Casimir scaling is a {\it good approximation}\ to the Yang-Mills
results. To be more quantitative, one could say that all lattice
results are within 10\% of the Casimir scaling prediction, and that
deviations from it are larger in four than they are in three
dimensions, in agreement with strong-coupling
predictions~\cite{DelDebbio:2001sj}. The taming of systematic errors
is a crucial matter for such lattice calculations, and it can only be
achieved by an intensive numerical analysis. In four dimensions, the
higher statistics simulations presented in
Ref.~\cite{DelDebbio:2001sj} show that corrections to the Casimir
scaling formula are statistically significant, and actually favor the
Sine scaling. Finally, it has been pointed out in
Ref.~\cite{DelDebbio:2003tk} that higher-dimensional representations
with common $N$-ality do yield the same string tension, as expected
because of gluon screening.

These numerical results trigger a few comments on Casimir scaling. The
original argument~\cite{Ambjorn:1984mb} was based on the idea that a
4-d gauge theory in a random magnetic field could be described by a
2-d theory without such a field. Besides the numerical results, there
is little support for such an argument in QCD; moreover it is not
clear that the same hypothesis could explain the approximate Casimir
scaling observed in three dimensions. On the other hand, Casimir
scaling appears ``naturally'' as the lowest order result, both at
strong-coupling in the case of $k$-strings in the hamiltonian
formulation of gauge theories, and in the case of the spectrum of
bound states in chiral models. Corrections can be computed in the
strong-coupling formulation and they turn out to be $\propto (d-2)/N$
- see e.g. Ref.~\cite{DelDebbio:2001sj} for a summary of results and
references. While we are aware that strong-coupling calculations are
not directly relevant to describe the physics of the continuum theory,
we think that it is nonetheless instructive to have some quantitative
analytic control within that framework. Last but not least, Casimir
scaling also appears at the lowest order in the stochastic model of
the QCD vacuum~\cite{xxx}. In view of these considerations, it is fair
to say that approximate Casimir scaling should be a prerequisite for
any model of confinement, that corrections should be expected, and
that these corrections are liable to yield further informations about
the non-perturbative dynamics of strong interactions. Moreover, it
would be very interesting to improve our understanding of some other
aspects of the $k$-string spectrum, like e.g. the origin of the Sine
scaling for non-supersymmetric theories, or the structure of the
corrections to this scaling form.

In order to get more insight in the dynamics underlying the $k$-string
spectrum, we compute in this paper the ratio $R(k,N)$ in the 3-d
SU($N$) Georgi-Glashow model. This model, in the $(N=2)$-case, is a
classic example~\cite{pol} of a theory, which allows for an analytic
description of confinement. The latter is due to the plasma of
point-like magnetic monopoles, which produce random magnetic fluxes
through the contour of the Wilson loop. In the weak-coupling regime of
the model (which is assumed henceforth), this plasma is dilute, and
the interaction between monopoles is Coulombic, being induced by the
dual-photon exchanges. Since the energy of a single monopole is a
quadratic function of its flux, it is energetically favorable for the
vacuum to support a configuration of two monopoles of unit charge (in
the units of the magnetic coupling constant, $g_m$), rather than a
single, doubly-charged monopole. Owing to this fact, monopoles of unit
charge dominate in the vacuum, whereas monopoles with higher charges
tend to dissociate into them.  Summing over the grand canonical
ensemble of monopoles of unit charge, interacting with each other by
the Coulomb law, one arrives at an effective low-energy theory, which
is a 3-d sine-Gordon theory of a dual photon. The latter acquires a
mass (visible upon the expansion of the cosine potential) by means of
the Debye screening in the Coulomb plasma. The appearance of this
(exponentially small) mass and, hence, of a finite (albeit
exponentially large) vacuum correlation length is crucial for the
generation of the fundamental string tension, i.e., for the
confinement of the external fundamental matter.  It is worth noting
that a physically important interpretation of these ideas in terms of
spontaneous breaking of magnetic $Z_2$ symmetry has been presented in
reviews~\cite{Alik} and refs. therein.

In the next section, we formalize this qualitative discussion and
describe the SU($N$)-generalization of the model, thereby introducing
the notations used throughout this paper. In section~3, we proceed
directly with the evaluation of the $k$-string tensions in the
SU($N$)-version of the model, first in the dilute-plasma
approximation, and then with the leading non-diluteness correction
taken into account. In section~4, we consider other possible
corrections, which stem from the finiteness of the Higgs-boson mass
and from the closed electric strings, present in the model. Finally,
in section~5, the main results of the paper are summarized, and
possible generalizations to the 4-d case are discussed.

\section{The model}

The Euclidean action of the 3-d Georgi-Glashow model reads~\cite{pol}

\begin{equation}
\label{SGG}
S=\int
d^3x\left[\frac{1}{4g^2}\left(F_{\mu\nu}^a\right)^2+\frac12
\left(D_\mu\Phi^a\right)^2+ 
\frac{\lambda}{4}\left(\left(\Phi^a\right)^2-\eta^2\right)^2\right],
\end{equation}
where the Higgs field $\Phi^a$ transforms by the adjoint
representation, i.e.,
$D_\mu\Phi^a\equiv\partial_\mu\Phi^a+\varepsilon^{abc}A_\mu^b\Phi^c$.
The weak-coupling regime $g^2\ll m_W$ parallels then the requirement
that $\eta$ should be large enough to ensure the spontaneous symmetry
breaking from SU(2) to U(1). At the perturbative level, the spectrum
of the model in the Higgs phase is made of a massless photon, two
heavy, charged $W$-bosons with mass $m_W=g\eta$, and a neutral Higgs
field with mass $m_H=\eta\sqrt{2\lambda}$.

What is however more important is the nonperturbative content of the
model, represented by the famous 't~Hooft-Polyakov
monopole~\cite{'tHooft:1974qc,Polyakov:ek}. It is a solution to the
classical equations of motion, which has the following Higgs- and
vector-field parts
\begin{itemize}
\item
$\Phi^a=\delta^{a3}u(r)$, $u(0)=0$, 
$u(r)\stackrel{r\to\infty}{\longrightarrow}\eta-\frac{{\rm e}^{-m_Hr}}{gr}$;
\item
$A_\mu^{1,2}(\vec x)\stackrel{r\to\infty}{\longrightarrow}
{\cal O}\left({\rm e}^{-m_Wr}\right)$,
$H_\mu\equiv\varepsilon_{\mu\nu\lambda}\partial_\nu A_\lambda^3=
\frac{x_\mu}{r^3}-4\pi\delta(x_1)\delta(x_2)
\theta(x_3)\delta_{\mu 3}$;
\item
as well as the following action $S_0=\frac{4\pi\epsilon}{\kappa}$. Here,
$\kappa\equiv g^2/m_W$ is the weak-coupling parameter, 
$\epsilon=\epsilon\left(m_H/m_W\right)$ is a certain monotonic, slowly
varying function, $\epsilon\ge 1$, $\epsilon(0)=1$ (BPS-limit)~\cite{bps},
$\epsilon(\infty)\simeq 1.787$~\cite{kirk}.
\end{itemize}

The following approximate saddle-point solution (which becomes exact
in the BPS-limit) has been found in ref.~\cite{pol}:

$$
S={\cal N}S_0+\frac{g_m^2}{8\pi}
\sum\limits_{{a,b=1\atop a\ne b}}^{\cal N}
\left(\frac{q_aq_b}{|\vec z_a-\vec z_b|}-
\frac{{\rm e}^{-m_H|\vec z_a-\vec z_b|}}{|\vec z_a-\vec z_b|}\right)
+ {\cal O}\left(g_m^2m_H{\rm e}^{-2m_H|\vec z_a-\vec z_b|}\right)
+ {\cal O}\left(\frac{1}{m_WR}\right),$$
where $m_W^{-1}\ll R\ll |\vec z_a-\vec z_b|$, $gg_m=4\pi$,
$[g_m]=[{\rm mass}]^{-1/2}$. Therefore, while at $m_H\to\infty$, the
usual compact-QED action is recovered, in the BPS-limit one has

$$S\simeq{\cal N}S_0+\frac{g_m^2}{8\pi}
\sum\limits_{{a,b=1\atop a\ne b}}^{\cal N}
\frac{q_aq_b-1}{|\vec z_a-\vec z_b|},$$ 
i.e., the interaction of two monopoles doubles for opposite and
vanishes for equal charges. Therefore, in this limit, the standard
monopole-antimonopole Coulomb plasma recombines itself into two
mutually noninteracting subsystems, consisting of monopoles and
antimonopoles. The interaction between the objects inside each of
these subsystems has a double strength with respect to the interaction
in the initial plasma.

When $m_H<\infty$, the summation over the grand canonical ensemble of
monopoles has been performed in ref.~\cite{dietz} and reads

$$
{\cal Z}_{\rm mon}=1+\sum\limits_{{\cal N}=1}^{\infty}
\frac{\zeta^{\cal N}}{{\cal N}!}\prod\limits_{a=1}^{\cal N}\int d^3z_a
\sum\limits_{q_a=\pm 1}^{}{\rm e}^{-S}=
$$

\begin{equation}
\label{ZMON}
=\int {\cal D}\chi{\cal D}\psi\exp\left\{-
\int d^3x\left[\frac12(\partial_\mu\chi)^2+\frac12(\partial_\mu\psi)^2
+\frac{m_H^2}{2}\psi^2-2\zeta{\rm e}^{g_m\psi}\cos(g_m\chi)\right]\right\}.
\end{equation}
Here, $\chi$ is the dual-photon field and $\psi$ is the field
additional with respect to compact QED, which describes the Higgs
boson. Furthermore, the monopole fugacity (i.e., the statistical
weight of a single monopole), $\zeta$, has the following
form~\cite{pol}

\begin{equation}
\label{0}
\zeta=\delta\frac{m_W^{7/2}}{g}
{\rm e}^{-S_0}.
\end{equation}
The function $\delta=\delta\left(m_H/m_W\right)$ is determined by the
loop corrections.  It is known~\cite{ks} that this function grows in
the vicinity of the origin (i.e., in the BPS limit). However, the
speed of this growth is such that it does not spoil the exponential
smallness of $\zeta$ in the weak-coupling regime under study.

For the analysis of $k$-strings in the next sections, we will need the
SU($N$)-generalization of the partition function~(\ref{ZMON}), which
reads

$$
{\cal Z}_{\rm mon}^N=
\int {\cal D}\vec\chi{\cal D}\psi\;\times
$$

\begin{equation}
\label{zMON}
\times\exp\left[-\int d^3x\left(\frac12\left(\partial_\mu\vec\chi\right)^2
+\frac12\left(\partial_\mu\psi\right)^2+\frac{m_H^2}{2}\psi^2
-2\zeta{\rm e}^{g_m\psi}
\sum\limits_{i=1}^{N(N-1)/2}
\cos\left(g_m\vec q_i\vec\chi\right)\right)\right].
\end{equation}
It has been taken into account here that, in the SU($N$)-case,
monopole charges are distributed along the $(N-1)$-dimensional
positive root vectors $\vec q_i$'s of the group SU($N$)~\cite{roots},
while charges of antimonopoles are represented by roots, that are
negative symmetric to those of monopoles.  Clearly, the dual-photon
field is now described by the $(N-1)$-dimensional vector $\vec\chi$.
In the next section, we will study $k$-strings in the compact-QED
limit of this theory, while corrections due to the propagation of the
Higgs field will be addressed in section~4.

\section{$k$-strings in the weakly coupled SU($N$) 3-d
  Georgi-Glashow model} 

In the compact-QED limit, the partition function~(\ref{zMON}) takes
the following sine-Gordon--type form:

\begin{equation}
\label{1}
{\cal Z}_{\rm mon}=
\int {\cal D}\vec\chi\exp\left[-\int d^3x\left(\frac12
\left(\partial_\mu\vec\chi\right)^2-2\zeta\sum\limits_{i=1}^{N(N-1)/2}
\cos\left(g_m\vec q_i\vec\chi\right)\right)\right].
\end{equation}
Owing to the orthonormality of roots,
$\sum\limits_{i=1}^{N(N-1)/2}q_i^\alpha
q_i^\beta=\frac{N}{2}\delta^{\alpha\beta}$,
$\alpha,\beta=1,\ldots,N-1$, the following value of the Debye mass of
the dual photon stems from eq.~(\ref{1}): $m=g_m\sqrt{N\zeta}$.

The $k$-string tension is defined by means of the $k$-th power of the
fundamental Wilson loop. The surface-dependent part of the latter is
contained in the following expression

\begin{equation}
\label{2}
\left\langle W_k({\cal C})\right\rangle_{\rm mon}=\sum\limits_{a_1,\ldots,
a_k=1}^{N}\left\langle\exp\left[-ig\left(
\sum\limits_{i=1}^{k}\vec\mu_{a_i}\right) \int d^3x\Sigma\left(\vec
x\right) \vec\chi\left(\vec x\right) \right]\right\rangle_{\rm mon},
\end{equation}
which is a consequence of the formula

$${\rm tr}{\,}\exp\left(i\vec {\cal O}\vec H\right)=
\sum\limits_{a=1}^{N}\exp\left(i\vec {\cal O}\vec\mu_a\right).$$ Here,
$\vec {\cal O}$ is an arbitrary $(N-1)$-component vector, $\vec H$ is
the vector of diagonal generators and $\vec\mu_a$, $a=1,\ldots, N$,
are the weight vectors of the fundamental representation of the group
SU($N$).  Furthermore, in eq.~(\ref{2})

$$\Sigma\left(\vec x\right)\equiv\int\limits_{\Sigma({\cal
C})}^{}d\sigma_\mu\left(\vec
x(\xi)\right)\partial_\mu^x\delta\left(\vec x- \vec x(\xi)\right),$$
where $\Sigma({\cal C})$ is an arbitrary surface bounded by the
contour ${\cal C}$ and parametrized by the vector $\vec x(\xi)$, with
$\xi=\left(\xi^1,\xi^2\right)$ standing for the 2-d coordinate,
$\xi\in[0,1]\times[0,1]$.

The independence of eq.~(\ref{2}) of the choice of $\Sigma({\cal C})$
can readily be seen in the same way as for the $(k=1)$-case (see
e.g. ref.~\cite{jh}).~\footnote{ It is a mere consequence of the
quantization condition $gg_m=4\pi$ and the fact that the product
$\vec\mu_a\vec q_i$ is equal either to $\pm\frac12$ or to $0$.} The
$\Sigma$-dependence rather appears in the weak-field, or low-density,
approximation, when one may keep only the quadratic term in the
expansion of the cosine in eq.~(\ref{1}). It has been shown in
ref.~\cite{jh} that the notion ``low-density'' implies that the
typical monopole density is related to the mean one, $\rho_{\rm
mean}=\zeta N(N-1)$, by the following sequence of inequalities:

$$ 
\rho_{\rm typical}\ll\zeta\cdot{\cal O}(N)\ll\rho_{\rm
mean}=\zeta\cdot{\cal O}\left(N^2\right).
$$ 
Below in this section, we will discuss in some more details the
correspondence between the low-density approximation and the large-$N$
one.

Denoting for brevity $\vec a\equiv-g\int d^3x\Sigma\left(\vec x\right)
\vec\chi\left(\vec x\right)$, we can rewrite eq.~(\ref{2}) as

\begin{equation}
\label{rep}
\left\langle W_k({\cal C})\right\rangle_{\rm mon}=\sum\limits_{{a_1,\ldots,
a_k=1}\choose{{\rm with}~ {\rm possible}~ {\rm
repetitions}}}^{N}\left\langle {\rm e}^{i\vec
a\left(\vec\mu_{a_1}+\cdots+\vec\mu_{a_k}\right)}\right\rangle,
\end{equation}
where in the low-density approximation the average is defined with
respect to the action

\begin{equation}
\label{quadr}
\int d^3x\left[
\frac12\left(\partial_\mu\vec\chi\right)^2+\frac{m^2}{2}\vec\chi^2\right].
\end{equation} 
Note that, in this approximation, the string tension for a given $k$
is the same for all surfaces $\Sigma({\cal C})$, which are large
enough in the sense $\sqrt{S}\gg m^{-1}$, where $S$ is the area of
$\Sigma({\cal C})$ (see the discussion in ref.~\cite{jh}). In
particular, the fundamental string tension reads~\cite{jh}
$\sigma_1=\frac{N-1}{2N}\bar\sigma$, where $\bar\sigma\equiv 4 \pi^2
\frac{\sqrt{\zeta N}}{g_m}$, and the factor $\frac{N-1}{2N}$ is the
square of a weight vector.

To evaluate eq.~(\ref{rep}) for $k>1$, we should calculate the
expressions of the form

\begin{equation}
\label{25}
\left(n\vec\mu_{a_i}+\sum\limits_{j=1}^{k-n}\vec\mu_{a_j}\right)^2,
\end{equation}
where $(k-n)$ weight vectors $\vec\mu_{a_j}$'s are mutually different
and also different from the vector $\vec\mu_{a_i}$. By virtue of the
formula
$\vec\mu_a\vec\mu_b=\frac12\left(\delta_{ab}-\frac{1}{N}\right)$, we
obtain for eq.~(\ref{25}):

\begin{equation}
\label{27}
\frac{N-1}{2N}\left(n^2+k-n\right)-\frac{1}{2N}\left[2n(k-n)+2\sum
\limits_{l=1}^{k-n-1}l\right]=   
\frac{k(N-k)}{2N}+\frac12\left(n^2-n\right).
\end{equation}
We should further calculate the number of times a term with a given
$n$ appears in the sum~(\ref{25}). In what follows, we will consider
the case $k<N$, although $k$ may be of the order of $N$. Then,
$C_k^n\equiv\frac{k!}{n!(k-n)!}$ possibilities exist to choose out of
$k$ weight vectors $n$ coinciding ones, whose index may acquire any
values from $1$ to $N$.  The index of any weight vector out of other
$(k-n)$ ones may then acquire only $(N-1)$ values, and so on. Finally,
the index of the last weight vector may acquire $(N-k+n)$
values. Therefore, the desired number of times, a term with a given
$n$ appears in the sum~(\ref{25}), reads:

$$
C_k^nN\cdot(k-n)(N-1)\cdot(k-n-1)(N-2)\cdots1(N-k+n)=
$$

\begin{equation}
\label{29}
=C_k^nA_N^{k-n+1}(k-n)!=\frac{k!N!}{n!(n+N-k-1)!},
\end{equation}
where $A_N^{k-n+1}\equiv\frac{N!}{(N-k+n-1)!}$. Equations~(\ref{27})
and (\ref{29}) together yield for the monopole contribution to the
Wilson loop, eq.~(\ref{rep}):

\begin{equation}
\label{299}
\left\langle W_k({\cal C})\right\rangle_{\rm mon}=k!N!{\rm e}^{-C\bar\sigma
  S}\cdot\sum\limits_{n=1}^{k} 
\frac{1}{n!(n+N-k-1)!}{\rm e}^{-\frac{n^2-n}{2}\bar\sigma S},
\end{equation}
where $C\equiv\frac{k(N-k)}{2N}$ is proportional to the Casimir of the
rank-$k$ antisymmetric representation of SU($N$). We have thus arrived
at a Feynman-Kac--type formula, where, in the asymptotic regime of
interest, $S\to\infty$, only the first term in the sum is
essential. The $k$-string tension therefore reads
$\sigma_k=C\bar\sigma$, that yields the Casimir-scaling law~(\ref{r}).
It is interesting to note that the Casimir of the original unbroken
SU($N$) group is recovered. This is a consequence of the Dirac
quantization condition~\cite{roots}, which distributes the quark
charges along the weights of the fundamental representation and the
monopole ones along the roots. The orthonormality of the roots then
yields the action (\ref{quadr}), which is diagonal in the dual
magnetic variables; the sum of the weights squared is responsible for
the Casimir factor, since
$C=\left(\vec\mu_{a_1}+\cdots+\vec\mu_{a_k}\right)^2$, where all $k$
weight vectors are different from each other. Therefore, terms where
all $k$ weight vectors are mutually different 
yield the dominant contribution to the sum~(\ref{rep}).  Their number
in the sum is equal to $\frac{k!N!}{(N-k)!}$, that corresponds to the
$(n=1)$-term in eq.~(\ref{299}).

Let us further address the leading correction to the obtained Casimir scaling, which
originates from the non-diluteness of plasma.  Expanding the cosine on
the r.h.s. of eq.~(\ref{1}) up to the quartic term, we obtain the
action~\footnote{One should use the formula

$$\sum\limits_{i=1}^{N(N-1)/2}q_i^\alpha q_i^\beta q_i^\gamma
q_i^\delta=\frac{N}{2(N+1)}
\left(\delta^{\alpha\beta}\delta^{\gamma\delta}+\delta^{\alpha\gamma}
\delta^{\beta\delta}+\delta^{\alpha\delta}\delta^{\beta\gamma} 
\right),$$ which stems from the orthonormality of roots.}

\begin{equation}
\label{quart}
\int d^3x\left[\frac12
\left(\partial_\mu\vec\chi\right)^2+\frac{m^2}{2}
\left(\vec\chi^2-\frac{g_m^2}{12(N+1)}\vec\chi^4\right)\right].
\end{equation}
By virtue of this formula, one can analyse the correspondence between
the $1/N$-expansion and corrections to the low-density approximation.
The natural choice for defining the behavior of the electric coupling
constant in the large-$N$ limit is the
QCD-inspired one, $g={\cal O}(N^{-1/2})$. To make some estimates in
this case, let us use an obvious argument that a $\vec\chi$-field
configuration dominating in the partition function is the one, where
every term in the action~(\ref{quart}) is of the order of unity. When
applied to the kinetic term, this demand tells us that the
characteristic wavelength $l$ of the field $\vec\chi$ is related to
the amplitude of this field as $l\sim|\vec\chi|^{-2}$. Substituting
further this estimate into the condition $l^3m^2|\vec\chi|^2\sim 1$,
we get $|\vec\chi|^2\sim m$. The ratio of the quartic and mass terms,
being of the order $|\vec\chi|^2g_m^2/N$, can then be estimated as
$\frac{mg_m^2}{N}=g_m^3\sqrt{\frac{\zeta}{N}}\sim N\sqrt{\zeta}$. With
the exponentially high accuracy, this ratio is small, provided
\begin{equation}
\label{upbo}
N\lesssim{\cal O}\left({\rm e}^{S_0/2}\right).
\end{equation} 
Therefore, the non-diluteness corrections are suppressed not at large
$N$, but rather at $N$'s bounded from above by a certain exponentially
large parameter.

To proceed with the study of the non-diluteness correction, one needs
to solve iteratively the saddle-point equation, corresponding to the
average~(\ref{2}) taken with respect to the approximate
action~(\ref{quart}). Since it has been demonstrated above that the
string tension is defined by the averages $\left\langle{\rm e}^{i\vec
a\left(\vec\mu_{a_1}+\cdots+\vec\mu_{a_k}\right)}\right\rangle$, where
all $k$ weight vectors are mutually different, let us restrict
ourselves only to such terms in the sum~(\ref{rep}).  Solving then the
saddle-point equation with the Ansatz
$\vec\chi=\vec\chi_0+\vec\chi_1$, where $|\vec\chi_1|\ll|\vec\chi_0|$,
we obtain for such a term:

\begin{equation}
\label{31}
-\ln\left\langle{\rm e}^{i\vec a
\left(\vec\mu_{a_1}+\cdots+\vec\mu_{a_k}\right)}\right\rangle=
\frac{g^2}{2}C\int d^3xd^3y\Sigma\left(\vec x\right)D_m
\left(\vec x-\vec y\right)
\Sigma\left(\vec y\right)+\Delta S,
\end{equation}
where $D_m\left(\vec x\right)={\rm e}^{-m|\vec x|}/(4\pi |\vec x|)$ is
the Yukawa propagator. The first term on the r.h.s. of eq.~(\ref{31})
yields the string tension $\sigma_k=C\bar\sigma$, while the second
term yields the desired correction. This term reads

$$
\Delta S=-\frac{2\pi^2}{3}\frac{(gmC)^2}{N+1}\int d^3x
\prod\limits_{l=1}^{4}\left[\int d^3x_lD_m
\left(\vec x-\vec x_l\right)\Sigma\left(\vec x_l\right)\right]=$$

\begin{equation}
\label{6}
=-\frac{\pi^2}{6}\frac{(gmC)^2}{N+1}
\int d\sigma_{\mu\nu}\left(\vec x_1\right)d\sigma_{\mu\nu}\left(\vec x_2\right)
d\sigma_{\lambda\rho}\left(\vec x_3\right)d\sigma_{\lambda\rho}
\left(\vec x_4\right)
\partial_\alpha^{x_1}\partial_\alpha^{x_2}\partial_\beta^{x_3}
\partial_\beta^{x_4}I,
\end{equation}
where $I\equiv\int d^3x\prod\limits_{l=1}^{4}D_m\left(\vec x-\vec
x_l\right)$.  The action~(\ref{6}) can be represented in the form

\begin{eqnarray}
\label{7}
\Delta S=&-&\frac{(gC)^2}{(N+1)m^5} \int d\sigma_{\mu\nu}(\vec x_1) 
d\sigma_{\mu\nu}(\vec x_2) D(\vec x_1 - \vec x_2) \; \times
\nonumber \\ 
&\times&\; \int d\sigma_{\mu\nu}(\vec x_3) 
d\sigma_{\mu\nu}(\vec x_4) D(\vec x_3 - \vec x_4) \;\times\; 
G(\vec x_1 - \vec x_3).
\end{eqnarray}
Here, $D$ and $G$ are some positive functions, which depend on $m|\vec
x_i-\vec x_j|$ and vanish exponentially at the distances $\gtrsim
m^{-1}$. They can be represented as $D(\vec x)=m^4{\cal D}(m|\vec
x|)$, $G(\vec x)=m^4{\cal G}(m|\vec x|)$, where the functions ${\cal
D}$ and ${\cal G}$ are dimensionless.~\footnote{Our investigations can
readily be translated to the Stochastic Vacuum Model of QCD~\cite{xxx}
for the evaluation of a correction to the string tension, produced by
the four-point irreducible average of field strengths. In that case,
the functions $D$ and $G$ would be proportional to the gluonic
condensate.}

The derivative expansion yields as leading terms the Nambu-Goto
actions:
\begin{equation}
\int d\sigma_{\mu\nu}(\vec x_1) 
d\sigma_{\mu\nu}(\vec x_2) D(\vec x_1 - \vec x_2) =
\sigma_D\int d^2\xi\sqrt{{\sf g}(\vec
x_1)}+{\cal O} \left(\sigma_D/m^2\right).
\end{equation}
Here, $\sigma_D=2m^2\int d^2z{\cal D}(|z|)$ (with $z$ being
dimensionless) and ${\sf g}(\vec x)=\det\| {\sf g}^{ab}(\vec x)\|$ is
the determinant of the induced-metric tensor ${\sf g}_{ab}(\vec x)=
(\partial_a \vec x)(\partial_b \vec x)$, where
$\partial_a\equiv\partial/\partial\xi^a$.  Next, the infinitesimal
world-sheet element $d\sigma_{\mu\nu}(\vec x)$ can be represented as
$d\sigma_{\mu\nu}(\vec x) =\sqrt{{\sf g}(\vec x)} t_{\mu\nu}(\vec
x)d^2\xi$, where $t_{\mu\nu}(\vec x)=\varepsilon^{ab}(\partial_a \vec
x)(\partial_b \vec x)/\sqrt{{\sf g}(\vec x)}$ is the
extrinsic-curvature tensor of $\Sigma$.  We may further take into
account that we are interested in the {\it leading} term of the
derivative expansion of the action $\Delta S$, which corresponds to
the so short distance $\left|\vec x_1-\vec x_3\right|$, that
$t_{\mu\nu}\left(\vec x_1\right)t_{\mu\nu}\left(\vec x_3\right) \simeq
2$. (Higher terms of the derivative expansion contain derivatives of
$t_{\mu\nu}$ and do not contribute to the string tension.) This yields
for the integral in eq.~(\ref{7}):
$$
\frac{\sigma_D^2}{2}
\int d\sigma_{\mu\nu}\left(\vec x_1\right)d\sigma_{\mu\nu}
\left(\vec x_3\right)
G\left(\vec x_1-\vec x_3\right)=
\frac{\sigma_D^2}{2} \left[\sigma_G\int d^2\xi\sqrt{\sf g}+{\cal O}
\left(\sigma_G/m^2\right)\right],$$ 
where $\sigma_G=2m^2\int d^2z{\cal G}(|z|)$. We finally obtain from
eq.~(\ref{7}):

$$
\Delta\sigma_k\simeq\frac{(gC)^2\sigma_D^2\sigma_G}{2(N+1)m^5}=
\frac{\alpha}{4}
\frac{(gC)^2m}{N+1}=\alpha\frac{\bar\sigma C^2}{N+1},$$
where $\alpha$ is some dimensionless positive constant. This yields

$$
\sigma_k+\Delta\sigma_k=\bar\sigma C\left(1+\frac{\alpha C}{N+1}\right).
$$ 
In the limit $k\sim N\gg 1$ of interest, the obtained correction is
${\cal O}(1)$, while for $k={\cal O}(1)$ it is obviously ${\cal
O}(1/N)$.  The latter fact enables one to write down the following
final result for the leading correction to eq.~(\ref{r}) due to the
non-diluteness of monopole plasma:

\begin{equation}
\label{ratio}
R(k,N)+\Delta R(k,N)\equiv
\frac{\sigma_k+\Delta\sigma_k}{\sigma_1+\Delta\sigma_1}=
C(k,N)\left[1+\alpha\frac{(k-1)(N-k-1)}{2N(N+1)}\right].
\end{equation}
This expression is as invariant under the replacement $k\to
N-k$ as the expression~(\ref{r}), which does not account for
non-diluteness. The fact that at $k\sim N\gg 1$ the obtained
correction to the Casimir-scaling law is ${\cal O}(1)$, indicates that
non-diluteness effects, once being taken into account, can
significantly distort the Casimir-scaling behavior.

\section{Corrections due to the Higgs field and closed strings}

As we have seen in section~2, when one deviates from the compact-QED
limit, i.e., makes the Higgs field not infinitely heavy, it starts
propagating and opens up an interaction in the monopole plasma via the
scalar (rather than vector) component of the monopole solution. The
respective partition function is given by eq.~(\ref{zMON}).  Averaging
in that equation over the Higgs field by means of the cumulant
expansion one gets in the second order of this expansion~\cite{mpla}:

$$
{\cal Z}_{\rm mon}\simeq
\int {\cal D}\vec\chi\exp\left[-
\int d^3x\left(\frac12(\partial_\mu\vec\chi)^2-2\xi\sum\limits_{i=1}^{N(N-1)/2}
\cos\left(g_m\vec q_i\vec\chi\right)\right)+\right.
$$

\begin{equation}
\label{morequart}
\left.+2\xi^2\int d^3xd^3y\sum\limits_{i,j=1}^{N(N-1)/2}
\cos\left(g_m\vec q_i\vec\chi(\vec x)\right){\cal K}(\vec x-\vec y)
\cos\left(g_m\vec q_j\vec\chi(\vec y)\right)\right].
\end{equation}
In this equation,
$\xi\equiv\zeta\exp\left[\frac{g_m^2}{2}D_{m_H}\left(m_W^{-1}\right)\right]$
is the modified fugacity (which can be shown to remain exponentially
small as long as the cumulant expansion is convergent) and ${\cal
K}({\bf x})\equiv {\rm e}^{g_m^2D_{m_H}({\bf x})}-1$.  The Debye mass
of the dual photon, stemming from eq.~(\ref{morequart}), reads

$$m=g_m\sqrt{N\xi}\left[1+\xi I\frac{N(N-1)}{2}\right],$$ where
$I\equiv\int d^3x{\cal K}({\bf x})$.  At $m_H\sim m_W$, the following
value of $I$ has been obtained~\cite{mpla}:

\begin{equation}
\label{i}
I\simeq\frac{4\pi}{m_Hm_W^2}\exp\left(\frac{4\pi}{\kappa}{\rm e}^{-m_H/m_W}\right).
\end{equation}
The parameter of the cumulant expansion is ${\cal O}\left(\xi
IN^2\right)$.  By virtue of eqs.~(\ref{0}) and~(\ref{i}), one can
readily see that the condition for this parameter to be
(exponentially) small reads
$N^2<\exp\left[\frac{4\pi}{\kappa}\left(\epsilon-\frac{1}{{\rm
e}}\right)\right]$.  Approximating $\epsilon$ by its value at
infinity, we find

\begin{equation}
\label{upbo1}
N<{\rm e}^{8.9/\kappa}.
\end{equation} 
Therefore, when one takes into account the propagation of the heavy
Higgs boson, the necessary condition for the convergence of the
cumulant expansion is that the number of colors may grow not
arbitrarily fast, but should rather be bounded from above by some
parameter, which is nevertheless exponentially large.  A similar
analysis can be performed in the BPS limit, $m_H\ll g^2$. There, one
readily finds $I\simeq\left(g_m/m_H\right)^2$, and

$$
\xi IN^2\propto N^2\exp\left[-\frac{4\pi}{\kappa}
\left(\epsilon-\frac12\right)\right].
$$ 
Approximating $\epsilon$ by its value at the origin, we see that the
upper bound for $N$ in this limit is smaller than in the vicinity of
the compact-QED limit and reads

\begin{equation}
\label{upbo2}
N<{\rm e}^{\pi/\kappa}.
\end{equation}
Note that both constraints~(\ref{upbo1}) and (\ref{upbo2}) are
more severe, in the respective limits, than the
constraint~(\ref{upbo}). Indeed, at $m_H\sim m_W$, eq.~(\ref{upbo})
reads [with the same accuracy as eq.~(\ref{upbo1})] $N\lesssim{\cal
O}\left({\rm e}^{11.2/\kappa}\right)$, and in the BPS limit it
obviously takes the form $N\lesssim{\cal O}\left({\rm
e}^{2\pi/\kappa}\right)$.

Upon the expansion of cosines in eq.~(\ref{morequart}), one can see
that the action~(\ref{quart}) becomes replaced by

$$\int d^3x\left[\frac12 \left(\partial_\mu\vec\chi\right)^2+
\frac{m^2}{2}\left(\vec\chi^2-\frac{g_m^2}{12(N+1)}\left(1+3\xi
IN(N+1)\right) \vec\chi^4\right)\right].$$ 
Following further the same steps which led from eq.~(\ref{quart}) to
eq.~(\ref{ratio}), we arrive at the following correction to the
latter:

\begin{equation}
\label{hi}
R(k,N)+\Delta R(k,N)
=C(k,N)\left\{1+\alpha\frac{(k-1)(N-k-1)}{2}
\left[\frac{1}{N(N+1)}+3\xi I\right]\right\}.
\end{equation}
The obtained correction is therefore exponentially small as long as
the parameter of the cumulant expansion is.

Another correction to eq.~(\ref{r}) is produced by closed $k$-strings.
Such strings are always present in the sector of a theory, where open
$k$-strings are present (see e.g. ref.~\cite{kk}). Owing to the
above-established fact that the open-string tension is saturated by
such configurations where all $k$ weight vectors are mutually
different, we may restrict ourselves to consideration of closed
strings of the same kind only. Then, according to eq.~(\ref{2}), the
statistical weight of the interaction of the dual photon with such a
closed string reads

\begin{equation}
\label{inter}
\exp\left[ig\left(\vec\mu_{a_1}+\cdots+\vec\mu_{a_k}\right)
\int d^3x\Sigma_\mu\partial_\mu\vec\chi\right],
\end{equation}
where $\Sigma_\mu(\vec x, \Sigma)\equiv \int\limits_{\Sigma}^{}
d\sigma_\mu\left(\vec x(\xi)\right)\delta\left(\vec x- \vec
x(\xi)\right)$ is the vorticity tensor current defined at the
closed-string world-sheet $\Sigma$. To model the grand canonical
ensemble of closed strings, one can proceed along the lines of
ref.~\cite{cor}. Namely, one can use the fact that these strings are
short-living (virtual) objects, whose typical sizes are much smaller
than the typical distances between them; therefore, similarly to
monopoles in the 3-d Georgi-Glashow model, closed strings form a
dilute plasma.  The vorticity tensor current of the ${\cal N}$-string
configuration, i.e. the ${\cal N}$-string density, (${\cal N}\ge 1$)
reads~\footnote{We also set by definition $\Sigma_\mu^{{\cal
N}=0}=0$.}  $\Sigma_\mu^{\cal N}=\sum\limits_{i=1}^{\cal
N}n_i\Sigma_\mu\left(\vec x, \Sigma_i\right)$, where $n_i$'s are
winding numbers. Similarly to monopoles, only strings with
minimal winding numbers, $n_i=\pm 1$, survive, whereas those with
$\left|n_i\right|>1$ dissociate into them. The summation over the
grand canonical ensemble of closed strings replaces then
eq.~(\ref{inter}) by

\begin{equation}
\label{grand}
\sum\limits_{{\cal N}=0}^{\infty}\frac{{\bar\zeta}^{\cal N}} {{\cal
N}!}\prod\limits_{i=0}^{\cal N}\int d^3y_i \sum\limits_{n_i=\pm
1}^{}\left\langle\exp\left[ig\left(\vec\mu_{a_1}+\cdots+
\vec\mu_{a_k}\right)\int d^3x \Sigma_\mu^{\cal
N}\partial_\mu\vec\chi\right]\right\rangle_{\vec z_i(\xi)}.
\end{equation}
Here, $\bar\zeta$ is the fugacity of a single closed string, which has
the dimensionality $[{\rm mass}]^3$, and is an exponentially small
quantity in the sense that the mean distance between neighbors in the
string plasma, ${\cal O}\left({\bar\zeta}^{-1/3}\right)$, is
exponentially large with respect to the characteristic string size.
Also, in eq.~(\ref{grand}), the world-sheet coordinate of the $i$-th
strings has been split as $\vec x_i(\xi)=\vec y_i+ \vec z_i(\xi)$,
where the vector $\vec y_i=\int d^2\xi\vec x_i(\xi)$ describes the
position of the string, whereas the vector $\vec z_i(\xi)$ describes
its shape. Furthermore, the translation- and $O(3)$ invariant measure
of integration over string shapes has been denoted as
$\left\langle\ldots\right\rangle_{\vec z_i(\xi)}$. The final
expression for the average does not depend on a particular form of
this measure. (The only thing which matters is the normalization
$\left\langle 1\right\rangle_{\vec z_i(\xi)}=1$.) The result of the
average rather does depend on the choice of the UV cutoff, which in
the theory under study is, however, unambiguous: $\Lambda=m_W$.

Equation~(\ref{grand}) produces then the following factor to the 
r.h.s. of eq.~(\ref{1}):

\begin{equation}
\label{str}
\exp\left\{2\bar\zeta\int d^3x\cos\left[\frac{g\left(\vec\mu_{a_1}+
\cdots+\vec\mu_{a_k}\right)\left|\partial_\mu\vec\chi\right|}{m_W^2}
\right]\right\},
\end{equation}
where $\int d^3x$ is nothing else, but the integration over the string
position. Also, in eq.~(\ref{str}), the absolute value is taken with
respect to the Lorentz indices only~\footnote{I.e.,
$\left|\partial_\mu\vec\chi\right|^\alpha=
\sqrt{\sum\limits_{\mu=1}^{3}(\partial_\mu\chi^\alpha)
(\partial_\mu\chi^\alpha)}$.}.
In the case of a dilute string plasma under study, we may approximate
the cosine in eq.~(\ref{str}) by the quadratic term only, that yields
the following addendum to the action~(\ref{quadr}):

$$
\frac{\bar\zeta}{m_W^3}\kappa\int d^3x\left[\left(\vec\mu_{a_1}+
\cdots+\vec\mu_{a_k}\right)\partial_\mu\vec\chi\right]^2.
$$ 
To evaluate with respect to this modified action the
average~(\ref{rep}) (approximated again by a single term where all
weight vectors are mutually different), we solve the respective
saddle-point equation with the natural Ansatz
$\vec\chi=\left(\vec\mu_{a_1}+\cdots+\vec\mu_{a_k}\right)\chi$. The
saddle-point equation,

$$
\left[-(1+\beta)\partial^2+m^2\right]\chi=-ig\Sigma,~~ \beta\equiv
2\kappa C\frac{\bar\zeta}{m_W^3},
$$
due to the smallness of the parameter $\beta$, has an approximate solution

$$ \chi^{\rm s.p.}(\vec x)\simeq -ig(1-\beta)\int d^3yD_{m_{*}} (\vec
x-\vec y)\Sigma(\vec y),$$ 
where $m_{*}=m\left(1-\frac{\beta}{2}\right)$ is the shifted Debye
mass. (This shift can be interpreted as an ``antiscreening'' of the
dual photon by closed strings, which carry an {\it electric} flux and
therefore diminish the effect of screening by {\it magnetic}
monopoles.) The string tension accordingly reads [cf. the
Casimir-scaling expression $\sigma_k=C\bar\sigma$]
$\sigma_k=C\left(1-\frac{3\beta}{2}\right)\bar\sigma$, so that
eq.~(\ref{r}) becomes modified as follows:

\begin{equation}
\label{RR}
R(k,N)+\Delta R(k,N)=C(k,N)\left[1-\frac{3\kappa}{2}
\frac{\bar\zeta}{m_W^3}\frac{(k-1)(N-k-1)}{N}\right].
\end{equation}
Although the numerator of the obtained correction is the same as that
of eq.~(\ref{ratio}), the denominator is obviously different in the
factor $(N+1)$. Consequently, although the correction due to closed
strings is small in the factor ${\cal
O}\left(\kappa\bar\zeta/m_W^3\right)$, at $k\sim N$ it scales as
${\cal O}\left(N^1\right)$, rather than ${\cal O}\left(N^0\right)$, as
the correction in eq.~(\ref{ratio}) does. Besides the $N$-dependence,
the obtained correction differs from that of eq.~(\ref{ratio}) by its
sign. However, eq.~(\ref{RR}) never becomes negative. Indeed, at
$k\sim N\gtrsim {\cal O}\left(\frac{m_W^3}{\kappa\bar\zeta}\right)$,
when this could have happened, the parameter $\beta$ becomes of the
order of unity, i.e. the original approximation $\beta\ll 1$, under
which eq.~(\ref{RR}) has been derived, breaks down. Therefore, at such
values of $k$ and $N$, eq.~(\ref{RR}) becomes merely invalid.

\section{Summary and discussion}
In the present paper, we have explored $k$-string tensions in the
SU($N$) 3-d Georgi-Glashow model.  The advantage of this model for
such an analysis is that confinement holds in it at weak coupling and
is therefore under a full analytic control. We have first addressed
the case of a very dilute monopole plasma and found there the
Casimir-scaling behavior for the $k$-string tension. After that, we
have proceeded with the leading non-diluteness correction and found
that, at $k\ll N$ it behaves as ${\cal O}(1/N)$, whereas at $k\sim N$
this correction is of the order of the leading term. The latter fact
means that, in the regime $k\sim N\gg 1$, the non-diluteness effects
can significantly change the Casimir-scaling behavior. The
non-diluteness correction increases the value of the ratio of string
tensions.

We have further addressed two other corrections to the Casimir
scaling. One of these is produced by the Higgs-mediated interaction of
monopoles, while the other is due to closed $k$-strings. The
Higgs-inspired effect appears as an addendum to the non-diluteness
correction and has the same sign as that correction (i.e., increases
the ratio of string tensions). However, the effect of the Higgs field
is small with respect to the non-diluteness correction, as long as the
cumulant expansion, one applies for the average over the Higgs field,
is convergent.  On the opposite, the correction to the Casimir
scaling, produced by closed strings, becomes significant at $k\sim
N\gg 1$.  Contrary to the non-diluteness and Higgs effects, this
correction diminishes the value of the Casimir-scaling ratio.

To conclude, notice that the obtained leading Casimir-scaling behavior
may, with a certain care, be translated to the SU($N$)-analogue of 4-d
compact QED. Obviously, this model can only be viewed as a continuum
limit of the respective lattice action~\cite{pol}.  Its crucial
difference from the above-discussed 3-d counterpart is that, in this
model, confinement holds at strong (rather than weak, as in 3-d)
coupling, and instead of the dilute plasma of point-like monopoles one
has a non-dilute ensemble of proliferating monopole loops (currents).
However, as we have argued in section~3, the non-diluteness
corrections to the quadratic dual-photon action are small, provided
$N$ is bounded from above by some exponentially large parameter and we
will exploit this fact below.  One may, nevertheless, take
non-diluteness into account by noting that it changes the interaction
between monopoles from Coulombic (as in 3-d) to a different type,
described by some unknown kernel. On general grounds, one should
accept that momenta, transferred by the dual photon in the
interactions between monopoles, are conserved. Therefore, the
above-mentioned kernel should be translation-invariant, i.e. the
interaction of two ${\cal N}$-monopole currents~\footnote{The magnetic
coupling constant, $g_m$, is obviously dimensionless in four
dimensions. Note also that we set by definition $\vec j_\mu^{{\cal
N}=0}(x)=0$.}, $\vec j_\mu^{\cal N}(x)= g_m\sum\limits_{a=1}^{\cal
N}\vec q_{i_a} \oint dx_\mu^a(\tau)\delta\left(x-x^a(\tau)\right)$,
should have the form $\vec j_\mu^{\cal N}(x)K(x-y)\vec j_\mu^{\cal
N}(y)$.

Further, to sum over the grand canonical ensemble of monopoles, let us
proceed in the same way as we did for closed strings in the preceeding
section. Namely, let us split $x_\mu^a(\tau)$ as
$x_\mu^a(\tau)=y_\mu^a+z_\mu^a(\tau)$, where
$y_\mu^a=\int\limits_{0}^{1}d\tau x_\mu^a(\tau)$ is the position of
the monopole trajectory, whereas the vector $z_\mu^a(\tau)$ describes
its shape. Let us also introduce the fugacity of a single-monopole
loop, $\zeta$, which has the dimensionality $[{\rm mass}]^4$,
$\zeta\propto {\rm e}^{-S_{\rm mon}}$. Here, the action of a single
$a$-th loop, obeying the estimate $S_{\rm mon}\propto
g_m^2\int\limits_{0}^{1}d\tau \sqrt{\left(\dot z_\mu^a\right)^2}$, is
assumed to be of the same order of magnitude for all loops. The
summation over the grand canonical ensemble then reads:

$$
\sum\limits_{{\cal N}=0}^{\infty}\frac{\zeta^{\cal N}}{{\cal N}!}
\prod\limits_{a=1}^{\cal N}\int d^4y_a
\sum\limits_{i_a=\pm 1,\ldots,\pm\frac{N(N-1)}{2}}^{}
\left\langle\exp\left(ig_m\vec q_{i_a}
\oint dz_\mu^a\vec\chi_\mu(x_a)\right)\right\rangle=$$

$$=\exp\left[2\zeta\sum\limits_{i=1}^{N(N-1)/2}\int d^4y
\left\langle\cos\left(g_m\vec q_i\oint dz_\mu\vec\chi_\mu(y+z)\right)
\right\rangle\right]=$$

\begin{equation}
\label{fin}
=\exp\left\{2\zeta\sum\limits_{i=1}^{N(N-1)/2}\int d^4y
\sum\limits_{k=1}^{\infty}\frac{(-1)^k}{2k!}\left\langle 
g_m\vec q_i\left[\oint dz_\mu \sum\limits_{n=0}^{\infty}
\frac{\left(z_\nu\partial_\nu^y\right)^n}{n!}\vec\chi_\mu(y)\right]
\right\rangle^{2k}\right\},
\end{equation}
where $\vec\chi_\mu$ is the dual-photon field, and
$\left\langle\cdots\right\rangle$ denotes some rotation- and
translation-invariant average over shapes of the loops,
$\left\{z_\mu^a\right\}$.  The term in the action with $k=1$, $n=0$
generates the Debye mass of the dual photon. Indeed, this term yields

$$
2\zeta\cdot\frac{N}{2}\cdot\frac{g_m^2}{2}
\left\langle \oint dz_\mu\oint dz'_\nu\right\rangle
\int d^4y\vec\chi_\mu(y)\vec\chi_\nu(y),
$$ 
and, due to the rotation and translation invariance of the measure
$\left\langle\ldots\right\rangle$, the average $\left\langle\oint
dz_\mu\oint dz'_\nu\right\rangle$ has the form $L\delta_{\mu\nu}$,
where $L$ is some parameter of dimensionality $[{\rm length}]^2$. The
Debye mass then reads $m=g_m\sqrt{2N\zeta L}$. In the leading
low-energy approximation, we may disregard terms with $n\ge 1$ in the
sum~(\ref{fin}). As for the terms with $k\ge 2$ (which account for
non-diluteness), as it has been discussed above, they are small and
can be disregarded at [cf. eq.~(\ref{upbo})]

\begin{equation}
\label{upbo3}
N\lesssim{\cal O}\left({\rm e}^{S_{\rm mon}/2}\right).
\end{equation} 
Therefore, the leading part of the dual-photon action reads 
$$\frac12\int
d^4x\vec\chi_\mu\left[K^{-1}(x)+m^2\right]\vec\chi_\mu,$$ where
$K^{-1}$ stands for the inverse operator. This action is the 4-d
analogue of the action~(\ref{quadr}), while, as it has already been
discussed, the rest of the derivation of eq.~(\ref{299}) is entirely
based on the properties of weights of the fundamental
representation.~\footnote{ The only thing which matters is the
generation of the string tension $\bar\sigma$ itself, which always
takes place as long as the dual-photon propagator, $\left[\tilde
K^{-1}(p)+m^2\right]^{-1}$, has a massive pole.  In the general case,
i.e. unless $K$ is adjusted in such a way that at small $p^2$, $\tilde
K^{-1}(p)=-m^2+\ldots$, this indeed holds true.} Therefore, at $N$
bounded from above according to the inequality~(\ref{upbo3}), one
obtains the Casimir scaling also within the leading low-energy
approximation to the SU($N$)-analogue of {\it four-dimensional}
compact QED.

\acknowledgments D.A. is grateful to the whole staff of the Physics
Department of the University of Pisa for cordial hospitality and to
INFN for the financial support.  The work of L.D.D. has been supported
by INFN and by MIUR-Progetto Teoria e fenomenologia delle particelle
elementari.

\end{document}